\documentstyle[12pt,epsfig]{article}

\def\beq{\begin{equation}}
\def\eq{\end{equation}}
\def\eeq{\end{equation}}

\def\centeron#1#2{{\setbox0=\hbox{#1}\setbox1=\hbox{#2}\ifdim
\wd1>\wd0\kern.5\wd1\kern-.5\wd0\fi
\copy0\kern-.5\wd0\kern-.5\wd1\copy1\ifdim\wd0>\wd1
\kern.5\wd0\kern-.5\wd1\fi}}
\def\ltap{\;\centeron{\raise.35ex\hbox{$<$}}{\lower.65ex\hbox{$\sim$}}\;}
\def\gtap{\;\centeron{\raise.35ex\hbox{$>$}}{\lower.65ex\hbox{$\sim$}}\;}
\def\gsim{\mathrel{\gtap}}
\def\lsim{\mathrel{\ltap}}

\def\vsup{\vspace{.2in}}

\def\jfig#1#2#3{
 \begin{figure}
 \centering
\epsfysize=3.2in
 \hspace*{0in}
 \epsffile{#2} \vspace{.1in}
 \caption{#3}
 \label{#1}
 \end{figure}}

\begin{document}

\begin{titlepage}

\begin{center}
\vspace*{-1cm}
\hfill SU-ITP-03-13 \\
\hfill hep-ph/0307004 \\
\vskip .6in
{\Large \bf Discretuum versus Continuum Dark Energy} \\
\vskip 0.4in
{\bf Savas Dimopoulos}~ and ~{\bf Scott Thomas}

\vskip 0.15in

{\em Department of Physics\\
Stanford University, Stanford, CA 94305}
\end{center}
\vskip 0.35in

\baselineskip=15pt

\begin{abstract}
The dark energy equation of state for theories with either
a discretuum or continuum distribution of vacua is investigated.
In the discretuum case the equation of state is constant
$w=p/\rho=-1$.
The continuum case may be realized by an action with
large wave function factor $Z$ for the dark energy modulus
and generic potential.
This form of the action is quantum mechanically stable
and does not lead to measurable long range
forces or violations of the equivalence principle.
In addition, it has a special property which may be
referred to as super-technical naturalness that results in a
one-parameter family of predictions for the cosmological
evolution of the dark energy equation of state as a function
of redshift $w=w(z)$.
The discretuum and continuum predictions will be tested
by future high precision measurements of the expansion
history of the universe.
Application of large $Z$-moduli to a predictive theory
of $Z$-inflation is also considered.
\end{abstract}
\end{titlepage}

\baselineskip=16pt

\newpage














\section{Multi-Vacua}

The fundamental theory of nature may possess
many
vacua, each with distinct physics.
For example, there are believed to be large
classes of (meta)stable (non)-supersymmetric string and M-theory
vacua \cite{bp,landscape}.
In a multi-vacua theory physical parameters depend on
moduli fields which varying between vacua.
There are a priori two classes of multi-vacua theories
which may be distinguished experimentally.
In the first, which has come to be know as the
discretuum, the moduli have a significant mass gap
or are discrete.
In this case the physical parameters in any given
vacuum are constant and time independent.
In the second class, which will be referred to as the continuum,
the moduli interpolate continuously between vacua.
If the full potential of the theory is moduli dependent,
each modulus value is not strictly an independent vacuum.
But if the moduli evolution is slow enough on
cosmologically relevant time scales, each value
can effectively be considered to be an independent vacuum.
In this case the physical parameters may vary
and in principle be time dependent.

In this paper we focus on the vacuum energy
and distinguish the observable properties of
cosmological dark energy in the discretuum and continuum cases.
The effects on other
physical parameters turn out to be insensitive
to the multi-vacua, as discussed below.
The continuum case may be realized by a 
modulus with large wave function factor $Z$
which in principle might arise from infrared physics.
This realization of the continuum is 
quantum mechanically stable and does not lead to observable
long range moduli forces or violations of the equivalence principle.
In addition, it has a special property we refer to as
super-technical naturalness, which in the simplest
case leads to a one-parameter family of predictions
for the evolution of the dark energy equation of state.
This form of moduli action is also applicable to theories
of inflation.

Cosmologies which allow
a distribution of vacuum energies 
through a very slowly evolving field have been considered 
previously in the context of very flat 
potentials \cite{banks,linde,qref}.
Large $Z$ Lagrangians which could stabilize moduli \cite{oldZstab}
or likewise allow a distribution
of vacuum energies \cite{oldZweinberg,oldZother}
have also been considered.
Here we address properties of the discretuum and 
continuum including 
quantum stability, (super)-technical naturalness,
absence of long range forces, implications for 
evolution of the dark energy equation of state, and effects
on other physical parameters.




\section{Discretuum versus Continuum}

For a discretuum multi-vacua, the vacuum energy in a given vacuum
is by definition constant, $V=V_0$.
The magnitude 
is not predicted, but with a
large enough number of vacua there may be many 
which happen to have
an energy density consistent with cosmological observations.
However,
the density and pressure are related in the discretuum case
to the vacuum energy by $\rho = V_0$ and $p=-V_0$.
So the discretuum leads to the prediction that
the dark energy equation of state parameter is constant
\beq
w=p/\rho=-1
\eq
independent of redshift.

For a continuum multi-vacua, the vacuum energy depends continuously
on some moduli fields.
Allowing for possible space-time dependence of the moduli fields,
the relevant Lagrangian describing the dark energy is then
\beq
{1 \over 2} Z ~\partial_{\mu} \phi \partial^{\mu} \phi - V(\phi)
\label{generalS}
\eeq
where $\phi$ parameterizes the trajectory through moduli space,
$Z$ is the kinetic term wave function, and $V(\phi)$
is the modulus dependent full quantum potential of the entire theory.
Spatial gradients in the moduli fields are redshifted to
insignificant levels by an early phase of inflation.
Keeping only possible time dependence, the
equation of motion for the dark energy modulus from the Lagrangian
(\ref{generalS}) in an expanding background is then
\beq
Z \ddot{\phi} + 3 Z H \dot{\phi} -
  V^{\prime}(\phi) =0
\label{eom}
\eq
where $H=\dot{a}/a$ is the Hubble constant,
$^{\prime} \equiv d/d \phi$, and
possible moduli dependence of $Z$ is neglected.
The first term in (\ref{eom}) is the moduli inertial kinetic term,
the second is damping due to expansion of the universe,
and the final term is forcing arising from the potential.
The continuum dark energy equation of state
parameter is
\beq
w_{\phi} = {p_{\phi} \over \rho_{\phi} } =
{ {1 \over 2} Z \dot{\phi}^2 - V(\phi) \over
       {1 \over 2} Z \dot{\phi}^2 + V(\phi)   }
\label{eos}
\eeq
With possible time dependence from the equation of motion (\ref{eom})
the continuum then allows for the possibility of a
time and therefore redshift dependent dark energy equation of state,
$w=w(z)$.

The specific 
evolution of the continuum equation of state
depends on the form the Lagrangian (\ref{generalS}).
Here we consider a very general form for the potential in which
the moduli can range over values of order the Planck scale,
and allow for the possibility that the overall magnitude of the
potential is smaller than the Planck scale
\beq
V(\phi) = m^2 M_p^2 f(\phi/M_p)
\label{generalV}
\eq
where $f$ is a generic dimensionless function with order unity
range and domain.
A small overall magnitude for the potential could arise, for example,
from supersymmetry which protects the potential from quantum corrections
above the supersymmetry breaking scale.
Another possibility is that the fundamental scale is smaller
than the four-dimensional Planck scale as occurs
in theories with extra dimensions.
In either case since the supersymmetry breaking and/or fundamental scale
can not be smaller than roughly TeV,
the overall magnitude is bounded by
$m^2 M_p^2 \gsim$ TeV$^4$ so that $m^2/M_p^2 \gsim 10^{-60}$.

Cosmological observations indicate that the universe has recently
entered a phase of accelerated expansion \cite{accdata}.
This implies the existence of a
dark energy which is vacuum dominated and
a sizeable fraction of critical density,
$V_0 \lsim 3 H_0^2 M_p^2$,
where $V_0=V(\phi_0)$ and $\phi_0$ is the present
value of the continuum dark energy modulus.
Since this is much smaller than the minimum magnitude
for the potential given above, $H_0^2 / M_p^2 \sim 10^{-120}$,
the most natural assumption
which does not require tuning of any 
Lagrangian (\ref{generalS}) parameters
is that the continuum potential
(\ref{generalV}) at $\phi_0$ happens to be near a simple zero
$f(\phi_0) \simeq 0$ \cite{zeronote}.

Vacuum domination of the dark energy implies
first that the kinetic energy is somewhat smaller than the
potential,
${1 \over 2} Z \dot{\phi}^2 \lsim V_0$, and second
that the inertial
term in the modulus
equation of motion (\ref{eom}) is small compared
with the friction and potential terms,
$\ddot{\phi} \lsim 3 H \dot {\phi}$, so that the
equation of motion reduces to
$3 Z H \dot{\phi} \simeq V^{\prime}(\phi)$.
These conditions, along with the magnitude of the dark energy
density given above, for any potential $V(\phi)$ may be written
\beq
{1 \over Z^{n/2} }
 \left| { d^n V(\phi_0) \over d \phi^n } \right|
  \lsim H_0^2 M_p^{2-n}
\label{Vnconditions}
\eq
for $n=0,1,2$.
These are analogous to the slow roll conditions of inflation.
For a generic potential if the conditions on the first
and/or second derivative are close to
being saturated then vacuum domination over of order a Hubble
time requires the
conditions (\ref{Vnconditions}) to hold for all derivatives,
$n \geq 0$.
In terms of the dimensionless function in the potential
(\ref{generalV}) the conditions (\ref{Vnconditions}) are
\beq
{ d^n f(x_0) \over dx^n} \lsim Z^{n/2} ~{H_0^2 \over m^2}
\label{fconditions}
\eq
where 
$x=\phi/M_p$.
So although the smallness of the overall vacuum energy
may arise naturally from a simple zero of the modulus potential,
continuum dark energy requires at least one small dimensionless parameter,
characterized by
$H_0^2/m^2 \ll 1$,
in the modulus Lagrangian (\ref{generalS}).




\section{Kinetic Seizing}

For a canonically normalized modulus, $Z=1$,
the conditions (\ref{fconditions}) require
possibly a large number of small parameters
in order to obtain vacuum dominated continuum dark energy.
In this language these conditions may be stated as the requirements
that the potential be very flat, very low curvature, and very smooth.

However, in a general basis with wave function $Z$, all the conditions
(\ref{fconditions}) may be satisfied for sufficiently large $Z$
even with a generic 
function $f$ without any small parameters.
The $0$-th 
derivative condition is satisfied automatically
near a zero 
$f(x_0) \lsim H_0^2/m^2$.
The first derivative condition is satisfied for 
order one 
$f^{\prime}(x_0)$ if $Z^{1/2} \gsim m^2/H_0^2$.
The second and higher order derivative conditions (\ref{fconditions})
are then automatically satisfied by additional
powers of $Z^{(n-1)/2}$
for order one dimensionless derivatives.
The required wave function factor ranges from $Z^{1/2} \gsim 10^{60}$
for $m^2 M_p^2 \sim$ TeV$^4$ to
$Z^{1/2} \gsim 10^{120}$
for $m^2 M_p^2 \sim M_p^4$.

This description is of course formally equivalent to the
canonically normalized one by a field redefinition,
$\tilde{\phi} = Z^{1/2} \phi$.
But in this language it is clear that only a {\it single} small
parameter, $Z^{-1/2}$, is required to obtain continuum dark
energy.
And as discussed below, the large $Z$ form of the Lagrangian is
super-technically natural which is
more restrictive than the most general technically natural
Lagrangian which could give rise to vacuum dominated continuum
dark energy.
Also, vacuum domination 
in this language arises not from special flatness properties
of the potential but from the large kinetic inertia
and Hubble 
damping of the continuum modulus.
This seizes its evolution, as is apparent in the
vacuum dominated equation of motion,
$\dot{\phi} \simeq V^{\prime}(\phi)/(3 Z H)$.
In the large $Z$ limit, space-time gradients in the modulus field
are suppressed by a large cost in Lagrangian density (\ref{generalS}).
In the $Z \rightarrow \infty$ limit the modulus completely
seizes and effectively becomes a parameter.

The most striking feature of continuum dark energy
is the specific form of the modulus evolution for large $Z$.
The potential may be expanded about the present value
of the dark energy modulus
\beq
V(\phi) = \sum_{n=0}^{\infty}  m^2 M_p^{2-n}
 {1 \over n!} {d^n f(x_0) \over dx^n}
 ( \phi - \phi_0)^n
\label{potexp}
\eq
Keeping for the moment
only the first derivative term, the modulus total change
over the history of the universe is
\beq
\Delta \phi = \int dt ~ \dot{\phi} \sim
{V^{\prime} \over  3 Z H_0^2 } =
{m^2 M_p ~f^{\prime} \over 3 Z H_0^2 }
\label{Deltaphi}
\eq
Using this, the contribution of the $n$-th derivative
term in the expansion (\ref{potexp}) to the total fractional
change in the potential is
\beq
{\Delta V^{(n)} \over V_0} \sim
 {m^2 \over H_0^2} \left( { m^2 f^{\prime} \over Z H_0^2 }\right)^n
 { d^n f  \over dx^n} \lsim
 {1 \over Z^{n/2}} {m^2 \over H_0^2}~ { d^n f  \over dx^n}
\label{DeltaV}
\eq
where for the inequality the first derivative
vacuum domination condition
(\ref{fconditions}) has been used.
Now since only the first derivative condition (\ref{fconditions})
is restrictive for large $Z$, implying $Z^{1/2} \gsim m^2/H_0^2$,
the contributions to (\ref{DeltaV}) of
derivative terms beyond linear order
are suppressed by additional
powers of $Z^{(n-1)/2}$ and therefore have a negligible
effect on the evolution of the dark energy modulus
over any observable cosmological epoch.
This occurs because with seized evolution the
modulus does not make large enough excursions for terms
beyond the linear one in (\ref{potexp}) to be important.
And with a linear potential only the ratio $V^{\prime}/Z$ appears in the
equation of motion (\ref{eom}).
So for a given dark energy density today,
the evolution of the continuum dark energy modulus is then characterized
by a {\it single} dimensionless seizing parameter which may be taken
to be $Z^{-1}|M_p V^{\prime}/V_0|$.

In order to be at least technically natural, the
form of the Lagrangian (\ref{generalS}) with large $Z$ and
potential (\ref{generalV}) must be stable against
quantum corrections.
Integrating out matter fields
to which the moduli may couple does not spoil the form of
the potential (\ref{generalV}) with a generic order
one function $f$ and Planck scale cutoff.
In addition, since the continuum modulus propagator
involves the inverse wave function 
\beq
\langle 0| \phi(0) \phi(p) |0 \rangle =
  {Z^{-1}  \over p^2 - V^{\prime \prime}(\phi_0) }
\label{prop}
\eq
corrections from moduli self interactions also do not
modify the form of the potential.
In fact, with a Planck scale cutoff
the potential would be quantum mechanically stable against
self coupling corrections and
technically natural even if the derivative terms in the
expansion (\ref{potexp}) where larger by factors of
$Z^{(n-1)/2}$ for $n \geq 2$.
So the form of the Lagrangian (\ref{generalS}) with
potential (\ref{generalV}) is actually more restrictive
than required by technical naturalness.
We will refer to the property of the Lagrangian (\ref{generalS})
with self interactions (\ref{potexp}) which are smaller
by factors of $Z^{(n-1)/2}$ than
that required solely by technical naturalness as
super-technical naturalness \cite{supernote}.
It is super-technical naturalness which is responsible
for the negligible effects of higher order derivative
terms 
(\ref{potexp}) on the cosmological
evolution of the dark energy modulus.
It is important to note that super-technical naturalness
follows from a single small parameter, $Z^{-1}$,
and is therefore not an unnatural
tuning beyond technical naturalness.


Super-technical naturalness and stability against
details of ultraviolet physics is
enforced by an approximate $\phi \rightarrow \phi + C$
shift symmetry which is recovered in the $Z \rightarrow \infty$ limit.
This arises because at fixed space-time gradients
in this limit the potential
term becomes insignificant compared with the gradient terms.
Note that the wave function may vary over the full moduli
space of multi-vacua so that the shift symmetry need only be an
emergent approximate
symmetry in regions of large $Z$.



Finally, note that although the continuum modulus
is very light and may
couple to matter fields, it does not lead to
measurable long range forces or violations of the
equivalence principle because of the highly
seized propagation,
as evidenced by the inverse wave function in the
propagator (\ref{prop}).

The features of continuum dark energy presented above
are to be contrasted with quintessence models \cite{qref}.
Most models of this type require multiple small parameters
which are not protected by an approximate symmetry.
As such they generally are not stable quantum mechanically
or technically natural,
are sensitive to details of ultraviolet physics,
and require unnatural tuning of parameters.
In addition, without an approximate symmetry to protect
couplings, the light fields could lead to
violations of the equivalence principle \cite{longrange}.
Even with an approximate symmetry to protect couplings
and stabilize the model, the most general
technically natural potential dessribed above which leads to vacuum
dominated dark energy saturates all the derivative
conditions (\ref{Vnconditions}).
All the derivatives in the potential are then important
and the evolution of the equation of state is in general not
a priori predictable.
This is in contrast to continuum dark energy which
requires only one super-technically natural small parameter and in
the simplest version
enjoys a one-parameter family of predictions outlined below.




\section{Large $Z$}

The large wavefunctions required for continuum dark energy
are super-technically natural and could be natural if
they arise from some dynamics.
One possibility is that $Z$ depends on other moduli
which dynamically drive it towards large values.
Another possibility is that large $Z$ results from infrared
dynamics below the fundamental scale \cite{oldZweinberg}.
It remains an open challenge to develop mechanisms or models
which are both natural and super-technically natural and
realize such large $Z$.


The approximate shift symmetry $\phi \rightarrow \phi + C$
which emerges
for large $Z$ might seem susceptible to violations by
quantum gravity corrections at the fundamental scale.
Large $Z$ may also seem equivalent
to a trans-Plankian
range for the canonically normalized modulus
$\tilde{\phi} = Z^{1/2} \phi$ since the potential
(\ref{generalV}) is then a function of
$f(\tilde{\phi}/Z^{1/2}M_p)$.
However, if the dynamics which leads to a large $Z$ were
due to infrared physics below the fundamental scale,
the shift symmetry
would be an emergent infrared symmetry and
therefore immune from any quantum gravity corrections
or questions of trans-Planckian ranges.





\section{Continuum Dark Energy}

The cosmological evolution which results with continuum dark
energy depends on the modulus wave function $Z$ and slope
of the potential $V^{\prime}$.
In a general model $Z$ might depend on other moduli and
itself be time dependent.

Here for definiteness we make the minimal assumption that
$Z$ is constant.
In this case the evolution is specified in terms of the
single dimensionless seizing parameter
$Z^{-1}|M_p V^{\prime}/V_0|$ discussed above.
In order to determine the expansion history,
the coupled Friedman and modulus equations of motion (\ref{eom})
are integrated forward from a high redshift
in the matter dominated era with initial conditions chosen
to yield a given
$\Omega_{\phi} = ( {1 \over 2} Z \dot{\phi}_0^2 + V(\phi_0))/
(3 H_0^2 M_p^2)$.
The evolution of the scale factor is shown in Fig${.~1}$ for different
values of $Z^{-1}|M_p V^{\prime}/V_0|$.
For small values 
the modulus
motion is highly seized and once the universe is dark
energy dominated it enters a
de-Sitter phase of accelerated exponential expansion
for an extended period.
For any value of seizing, the dark energy modulus
eventually evolves to negative total energy density
at which time the Hubble constant changes sign
and the expansion reverses.
The $3ZH \dot{\phi}$ term in the moduli equation of motion
(\ref{eom}) becomes anti-damping in this epoch and pushes
the modulus to more negative values of the potential,
resulting in a rapid crunch \cite{wnote}.
\jfig{atplot}{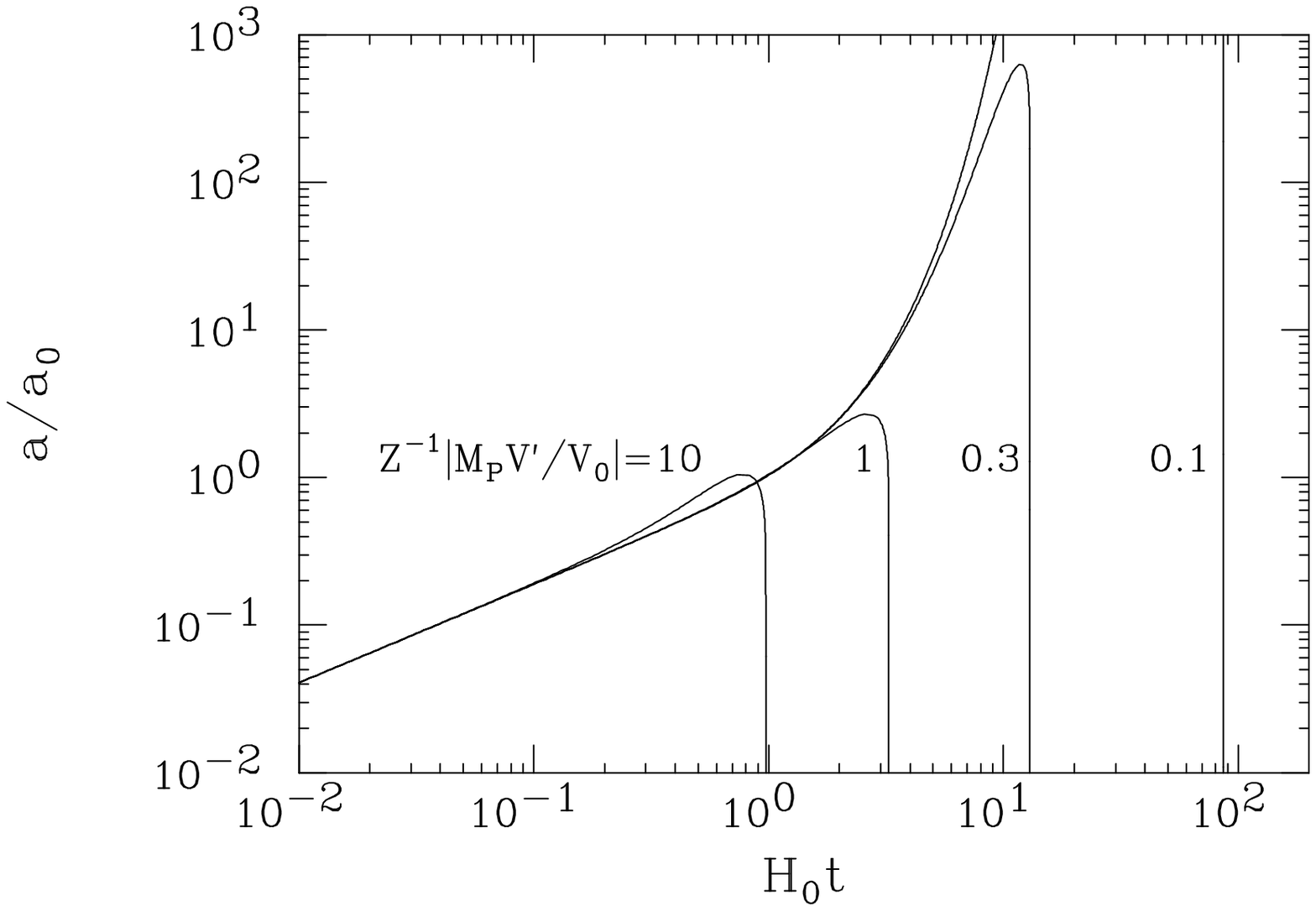}{Scale factor as a function of
time for different
continuum dark energy
seizing parameters
for $\Omega_{m}=0.3$ and $\Omega_{\phi}=0.7$.}
This feature is shared by any model which evolves to negative
energy density \cite{crunch}.
For $Z^{-1}|M_p V^{\prime}/V_0|$ of order unity the
de-Sitter phase is brief and the crunch
time is a few Hubble times from the beginning of
dark energy domination.

It may appear that the continuum modulus initial conditions
have to be rather specially chosen
in order to yield acceptable continuum dark energy.
However, there is a natural dynamical selection 
effect since only
regions of the universe (after inflation say)
which are near a zero of the continuum potential
can evolve to become large with extended eras of
radiation and matter domination followed much later
by dark energy domination.

Details of the transition from matter to dark energy domination
in general depend on the dark energy equation of state.
The continuum dark energy equations of state (\ref{eos})
as a function of redshift
for various values of constant seizing parameter
are shown in Fig${.~2}$.
At high redshift $w_{\phi} \rightarrow -1^+$ since
the larger Hubble constants there yield more effective
seizing and smaller modulus kinetic energy.
\jfig{eosplot}{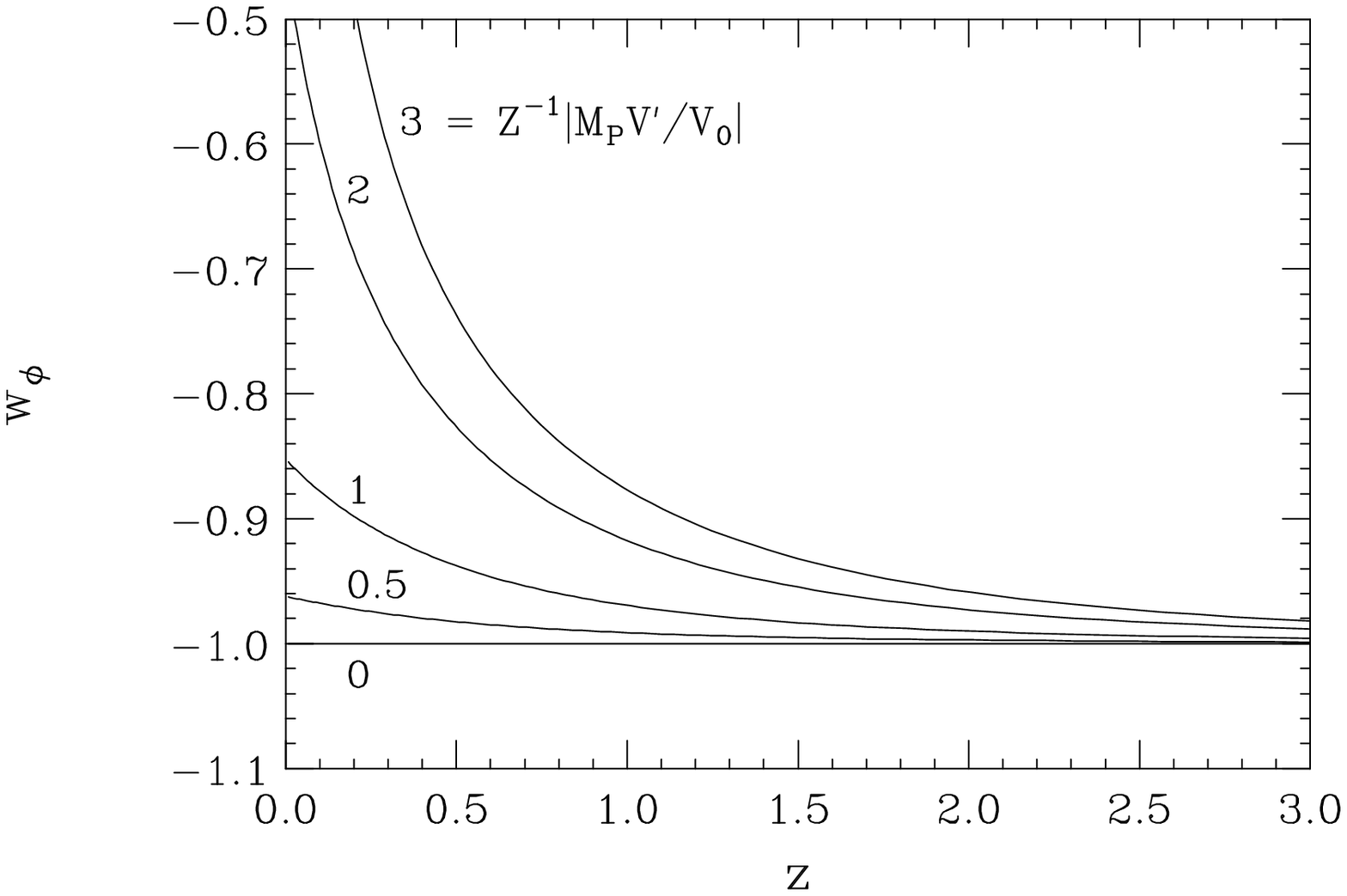}{Continuum dark energy equation of state
$w_{\phi}= p_{\phi} / \rho_{\phi}$ as a function of
redshift for different continuum 
seizing parameters for
$\Omega_{m}=0.3$ and $\Omega_{\phi}=0.7$.}
For $Z^{-1}|M_p V^{\prime}/V_0|$ of order unity
the vacuum domination condition (\ref{fconditions})
on the first derivative is just marginally satisfied,
and kinetic seizing is only marginally effective
at the transition from matter to dark energy domination
at a redshift $z \sim 1$.
In this case the continuum modulus begins to roll
at this transition epoch and the kinetic energy is not
too much smaller than the potential energy.
The dark energy equation of state can then differ
significantly from $w_{\phi}=-1$ at low redshift
and show considerable evolution.

The dark energy equation of state $w=w(z)$ is difficult to
extract directly from experimental measurements of the
expansion history.
It is more useful to consider experimental observables directly.
Fig${.~3}$ shows the magnitude--redshift relation
for various values of constant seizing parameter
relative to a constant vacuum energy cosmology.
This relation represents a one-parameter family of
predictions for the given cosmological parameters.
\jfig{mzplot}{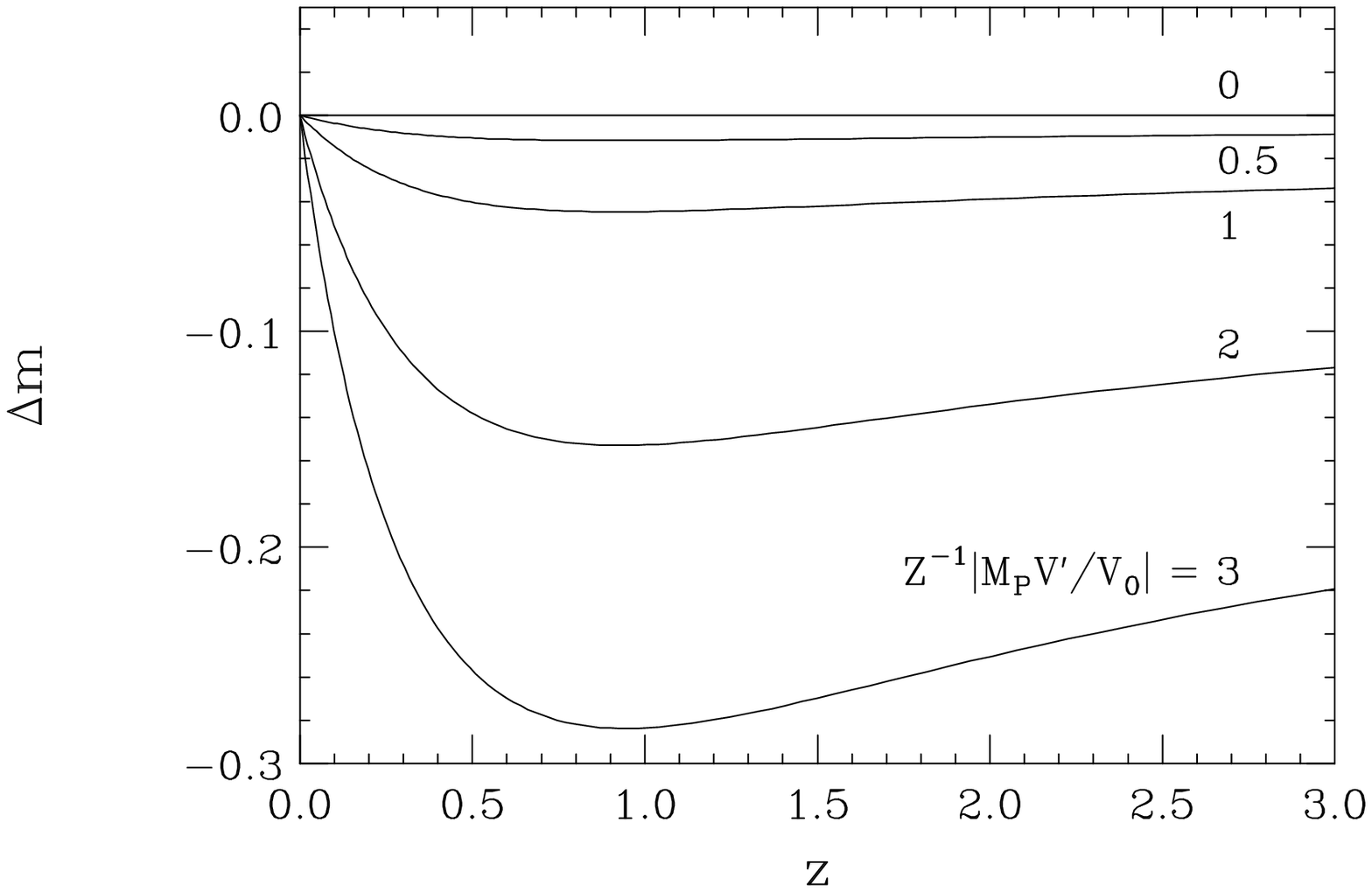}{Magnitude--redshift relation
for different
continuum seizing parameters
relative to constant vacuum energy
for $\Omega_{m}=0.3$ and $\Omega_{\phi}=0.7$.}
Future precision measurements of the magnitude--redshift relation
such as from the SuperNova Acceleration Probe (SNAP)
\cite{snap}
should be able to measure the seizing parameter at the
few \% level.
This, along with other measurements of the expansion
history from structure formation, weak lensing,
and the cosmic microwave background should eventually be able
to distinguish the continuum from other theories of dark
energy.
A detailed study of how well
such measurements
can test continuum dark energy \cite{numerics} in particular
by bounding higher derivatives in the dark energy
potential beyond linear order
will be presented elsewhere \cite{fmtoappear}.

Finally, it is worth considering what the expectation
might be for the value of $Z$.
In a multi-vacua theory the distribution of dark energy
modulus wave functions $Z$
over all the multi-vacua might be
a rapidly falling function at large $Z$.
This may in fact be likely since extremely large $Z$ seems
rather exceptional.
In this case the most likely value of $Z$ is
one not too much larger than the minimum value consistent
with the observed vacuum dominated dark energy.
This corresponds to a seizing parameter
$Z^{-1}|M_p V^{\prime}/V_0|$
not too much smaller than order unity.
So in this case the continuum dark energy modulus should
have begun to roll by the current epoch, leading to
a measurable evolution of the dark energy equation of
state \cite{linear}.

In any multi-vacua theory
the magnitude of the dark energy need not be
too much smaller than
the maximum allowed value \cite{weinberg}.
In a continuum multi-vacua theory we see that the
time dependent evolution
of the dark energy also need not be too much smaller
than the maximum allowed value.
The expectation that properties of the dark energy
are not much smaller than maximally acceptable
could be referred to as the principle of living dangerously,
since regions in which these properties are
significantly exceeded are unihabitable.




\section{Physical Parameters in the Multi-Vacua}

For a discretuum, all physical parameters, such as gauge
and Yukawa couplings, are constants and independent of time.
For a continuum, if these parameters depend on the continuum
moduli they are in principle time dependent.
However, from 
(\ref{Deltaphi}) the total
range of the continuum modulus over the history of the universe
is $\Delta \phi /M_p \lsim Z^{-1/2}$ where the inequality results from
the fist derivative vacuum domination condition (\ref{fconditions}).
So if physical parameters
are generic functions of the dark energy modulus with order unit range,
$g=g(\phi/M_p)$, then
the changes in these parameters,
$\Delta g \simeq g^{\prime} \Delta \phi / M_p
\lsim g^{\prime} Z^{-1/2}$ where
$ g^{\prime} \equiv dg(x_0)/dx$,
are unobservably small even for order one seizing
parameter.
So measurements of the
evolution of the dark energy equation of state appear to be unique
experimental probes of the nature of multi-vacua theory.




\section{$Z$-inflation}

Large $Z$ moduli
are also applicable to
inflation.
With only a moderate wave function factor, $Z^{1/2} \gsim 10$,
and a completely general potential,
seized evolution of a modulus can lead to an early
epoch of inflation with 
sufficient $e$-foldings to solve the horizon and flatness problems.
Since only the constant and linear parts of the potential
are important during seizing, the super-technically
natural feature of $Z$-inflation yields a predictive
relation between the spectral tilt, $n$,
derivative of the tilt, $dn/ d \ln k$, and tensor to scalar
ratio, ~~$T/S$ \cite{zinf}.
A large $Z$ inflaton has been considered
previously in the context of Brans-Dicke theory with exit from
inflation by bubble nucleation \cite{Zhyper} but without
regard to quantum stability,
or the predictive feature of super-technical naturalness.




\section{Conclusions}

Multi-vacua theories lead to distinct and testable predictions
for properties of cosmological dark energy.
In the discretuum case the dark energy equation of state
is constant $w=-1$.
In the continuum case super-technical naturalness of the
large $Z$ dark energy modulus leads to a one-parameter
family of predictions for the evolution of the dark energy
equation of state, $w=w(z)$, and associated
distance--redshift observables.
High precision measurements of the expansion history of the
universe will therefore provide important guidance
as to what type of fundamental theory may describe
our universe.


\vsup

We are grateful to A. Linde for several very valuable discussions, and 
also thank  N. Arkani-Hamed,
W. Hu, M. Kamionkowski, L. Susskind, and T. Tyson
for useful discussions.
This work was supported by
the US National Science Foundation under grant PHY02-44728,
the Alfred P. Sloan Foundation, Stanford
University, 
and was partially completed at the Aspen Center for
Physics.



\begin{thebibliography}{99}
\bibitem{zeronote}{A double zero of the potential,
$f(\phi_0) \simeq 0$ and $f^{\prime}(\phi_0) \simeq 0$
would require tuning of potential parameters and would not
be technically natural or quantum mechanically stable.}
\bibitem{supernote}{If large $Z$ 
arises from infrared effects then it is only
the low energy action which enjoys super-technical naturalness.}
\end{thebibliography}
\end{document}